\documentclass[aps,twocolumn,prd,superscriptaddress,preprintnumbers,showpacs]{revtex4}
\usepackage{graphicx}
\usepackage{bm}
\usepackage{amsmath}
\usepackage{amssymb}
\usepackage{amsthm}

\newcommand{\be}{\begin{equation}}
\newcommand{\ee}{\end{equation}}
\newcommand{\bea}{\begin{eqnarray}}
\newcommand{\eea}{\end{eqnarray}}
\newcommand{\bdm}{\begin{displaymath}}
\newcommand{\edm}{\end{displaymath}}
\newcommand{\beas}{\begin{eqnarray*}}
\newcommand{\eeas}{\end{eqnarray*}}

\begin{document}
\title{Probing unification scenarios with atomic clocks}

\author{M. C. Ferreira}
\email[]{up080302013@alunos.fc.up.pt}
\affiliation{Centro de Astrof\'{\i}sica, Universidade do Porto, Rua das Estrelas, 4150-762 Porto, Portugal}
\affiliation{Faculdade de Ci\^encias, Universidade do Porto, Rua do Campo Alegre, 4150-007 Porto, Portugal}
\author{M. D. Juli\~ao}  
\email[]{meinf12013@fe.up.pt}
\affiliation{Centro de Astrof\'{\i}sica, Universidade do Porto, Rua das Estrelas, 4150-762 Porto, Portugal}
\affiliation{Faculdade de Ci\^encias, Universidade do Porto, Rua do Campo Alegre, 4150-007 Porto, Portugal}
\affiliation{Faculdade de Engenharia, Universidade do Porto, Rua Dr Roberto Frias, 4200-465 Porto, Portugal}
\author{C. J. A. P. Martins}
\email[]{Carlos.Martins@astro.up.pt}
\affiliation{Centro de Astrof\'{\i}sica, Universidade do Porto, Rua das Estrelas, 4150-762 Porto, Portugal}
\author{A. M. R. V. L. Monteiro}
\email[]{up090322024@alunos.fc.up.pt}
\affiliation{Centro de Astrof\'{\i}sica, Universidade do Porto, Rua das Estrelas, 4150-762 Porto, Portugal}
\affiliation{Faculdade de Ci\^encias, Universidade do Porto, Rua do Campo Alegre, 4150-007 Porto, Portugal}
\affiliation{Department of Applied Physics, Delft University of Technology, P.O. Box 5046, 2600 GA Delft, The Netherlands}

\date{15 November 2012}

\begin{abstract}
We discuss the usage of measurements of the stability of nature's fundamental constants coming from comparisons between atomic clocks as a means to constrain coupled variations of these constants in a broad class of unification scenarios. After introducing the phenomenology of these models we provide updated constraints, based on a global analysis of the latest experimental results. We obtain null results for the proton-to-electron mass ratio ${\dot\mu}/{\mu}=(0.68\pm5.79)\times10^{-16}\, {\rm yr}{}^{-1}$ and for the gyromagnetic factor ${\dot g_p}/{g_p}=(-0.72\pm0.89)\times10^{-16}\, {\rm yr}{}^{-1}$ (both of these being at the $95\%$ confidence level). These results are compatible with theoretical expectations on unification scenarios, but much freedom exists due to the presence of a degeneracy direction in the relevant parameter space.
\end{abstract}

\pacs{12.10.Dm,06.20.Jr,06.30.Ft}
\maketitle

\section{\label{intro}Introduction} 

The observational evidence for the acceleration of the universe demonstrates that canonical theories of cosmology and particle physics are incomplete, if not incorrect. Several few-sigma hints of new physics exist, but so far these are smoke without a smoking gun; it's time to actively search for the gun.

The LHC evidence for the Higgs particle strongly suggests that fundamental scalar fields are among nature's building blocks. An obvious follow-up question is whether such fields also play a role in cosmology. They have been invoked to explain paradigms such as inflation, cosmological phase transitions or dynamical dark energy, but the most direct way to infer their presence is (arguably) to search for spacetime variations of nature's fundamental constants \cite{RSoc,GBerro,Uzan}. It is known that fundamental couplings run with energy, and many particle physics and cosmology models suggest that they should also roll with time. One example are cosmological models with dynamical scalar fields, including string theory.

Astrophysical measurements have led to claims for \cite{Webb,Reinhold,Dipole} and against \cite{Chand,King,Thompson} variations of the fine-structure constant $\alpha$ and the proton-to-electron mass ratio $\mu$ at redshifts $z\sim1-3$. An ongoing Large Programme at European Southern Observatory's Very Large Telescope should soon clarify matters, but a resolution may have to wait for a forthcoming generation of high-resolution ultra-stable spectrographs which include these measurements among their key science drivers. Answering this question can also shed light on the enigma of dark energy \cite{Parkinson,NunLid,Doran,Reconst}. 

Any Grand-Unified Theory predicts a specific relation between variations of $\alpha$ and $\mu$, and therefore  simultaneous measurements of both provide key consistency tests. Our work revisits the ideas of \cite{Coc,Luo} and applies them in the same spirit as \cite{Vieira,Angeles}, by using the most recent tests of the stability of fundamental constants using atomic clock comparisons to obtain direct constraints on the phenomenological parameters characterizing these unification scenarios.

\section{\label{unify}Phenomenology of unification}

We wish to describe phenomenologically a class of models with simultaneous variations of several fundamental couplings, such as the fine-structure constant $\alpha=e^2/\hbar c$, the proton-to-electron mass ratio $\mu=m_p/m_e$ and the proton gyromagnetic ratio $g_p$. The simplest way to do this is to relate the various changes to those of a particular dimensionless coupling, typically $\alpha$. Then if $\alpha=\alpha_0(1+\delta_\alpha)$ and
\begin{equation}
\frac{\Delta X}{X}=k_X\frac{\Delta\alpha}{\alpha}
\end{equation}
we have $X=X_0(1+k_X\delta_\alpha)$
and so forth.

The relations between the couplings will be model-dependent. In this section we follow \cite{Coc,Luo}, considering a class of grand unification models in which the weak scale is determined by dimensional transmutation and further assuming that relative variation of all the Yukawa couplings is the same. Finally we assume that the variation of the couplings is driven by a dilaton-type scalar field (as in \cite{Campbell}). With these assumptions one finds that the variations of $\mu$ and $\alpha$ are related through
\begin{equation}
\frac{\Delta\mu}{\mu}=[0.8R-0.3(1+S)]\frac{\Delta\alpha}{\alpha}\,,
\end{equation}
where $R$ and $S$ can be taken as free phenomenological (model-dependent) parameters. Their absolute value can be anything from order unity to several hundreds, although physically one usually expects them to be positive. (Nevertheless, for our present purposes they can be taken as free parameters to be constrained by data.)

For our purposes it's natural to assume that particle masses and the QCD scale vary, while the Planck mass is fixed. We then have
\begin{equation}
\frac{\Delta m_e}{m_e}=\frac{1}{2}(1+S)\frac{\Delta\alpha}{\alpha}
\end{equation}
(since the mass is simply the product of the Higgs VEV and the corresponding Yukawa coupling) and
\begin{equation}
\frac{\Delta m_p}{m_p}=[0.8R+0.2(1+S)]\frac{\Delta\alpha}{\alpha}\,.
\end{equation}
The latter equation is the more model-dependent one, as it requires modeling of the proton. At a phenomenological level, the choice $S=-1$, $R=0$ can also describe the limiting case where $\alpha$ varies but the masses don't. Further useful relations can be obtained \cite{flambaum1,flambaum2,flambaum3} for the g-factors for the proton and neutron
\begin{equation}
\frac{\Delta g_p}{g_p}=[0.10R-0.04(1+S)]\frac{\Delta\alpha}{\alpha}\,
\end{equation}
\begin{equation}
\frac{\Delta g_n}{g_n}=[0.12R-0.05(1+S)]\frac{\Delta\alpha}{\alpha}\,.
\end{equation}

These allow us to transform any measurement of a combination of constants (for example $\alpha$, $\mu$ and $g_p$) into a constraint on the $(R,S,\alpha)$ parameter space. For atomic clocks, the relevant g-factors are those for Rubidium and Caesium, so these need to be related to those of the nucleons. The way to do this stems from \cite{Luo,flambaum1,flambaum2,flambaum3}. Using a simple shell model one has 
\begin{equation}\label{gyro1a}
\frac{\Delta g_{Rb}}{g_{Rb}}\simeq0.736\frac{\Delta g_p}{g_p}\simeq[0.07R-0.03(1+S)]\frac{\Delta\alpha}{\alpha}\,
\end{equation}
\begin{equation}\label{gyro1b}
\frac{\Delta g_{Cs}}{g_{Cs}}\simeq-1.266\frac{\Delta g_p}{g_p}\simeq[-0.13R+0.05(1+S)]\frac{\Delta\alpha}{\alpha}\,;
\end{equation}
for our purposes in the following section, the following derived relation is also useful
\begin{equation}
\frac{\Delta (g_{Cs}/g_{Rb})}{(g_{Cs}/g_{Rb})}\simeq1.58\frac{\Delta g_{Cs}}{g_{Cs}}\simeq-2\frac{\Delta g_p}{g_p}\,.
\end{equation}
A more accurate phenomenological description (motivated from experimental results and including a dependence on $g_n$ and the spin-spin interaction) leads to 
\begin{equation}\label{gyro2a}
\frac{\Delta g_{Rb}}{g_{Rb}}\simeq[0.014R-0.007(1+S)]\frac{\Delta\alpha}{\alpha}\,
\end{equation}
\begin{equation}\label{gyro2b}
\frac{\Delta g_{Cs}}{g_{Cs}}\simeq[-0.007R+0.004(1+S)]\frac{\Delta\alpha}{\alpha}\,.
\end{equation}
Notice that these coefficients are very small, particularly in the last parametrization. 

\begin{table}
\begin{tabular}{|c|c|c|c|}
\hline
 Clock & $\nu_{AB}$ & ${\dot \nu_{AB}}/{\nu_{AB}}$ (yr${}^{-1}$) & Ref. \\ 
 \hline
 \hline
Hg-Al & $\alpha^{-3.208}$ & $(5.3\pm7.9)\times10^{-17}$ & \protect\cite{Rosenband} \\
\hline
Cs-SF${}_6$ & $g_{Cs}\mu^{1/2}\alpha^{2.83}$ & $(-1.9\pm0.12_{sta}\pm2.7_{sys})\times10^{-14}$ & \protect\cite{Shelkovnikov} \\
Cs-H & $g_{Cs}\mu\alpha^{2.83}$ & $(3.2\pm6.3)\times10^{-15}$ & \protect\cite{Fischer} \\
Cs-Sr & $g_{Cs}\mu\alpha^{2.77}$ & $(1.0\pm1.8)\times10^{-15}$ & \protect\cite{Blatt} \\
Cs-Hg & $g_{Cs}\mu\alpha^{6.03}$ & $(-3.7\pm3.9)\times10^{-16}$ & \protect\cite{Fortier} \\
\hline
Cs-Yb & $g_{Cs}\mu\alpha^{1.93}$ & $(0.78\pm1.40)\times10^{-15}$ & \protect\cite{Peik1} \\
Cs-Rb & ($g_{Cs}/g_{Rb})\alpha^{0.49}$ & $(0.5\pm5.3)\times10^{-16}$ & \protect\cite{Bize1} \\
\hline
Cs-Yb & $g_{Cs}\mu\alpha^{1.93}$ & $(0.49\pm0.41)\times10^{-15}$ & \protect\cite{Peik2} \\
Cs-Rb & ($g_{Cs}/g_{Rb})\alpha^{0.49}$ & $(1.39\pm0.91)\times10^{-16}$ & \protect\cite{Bize2} \\
\hline
\end{tabular}
\caption{\label{table1} Atomic clock constraints of varying fundamental couplings. The second column shows the combination of couplings to which the clock is sensitive, and the third column shows the corresponding experimental bound. The measurements in the first seven lines were the ones used in \protect\cite{Luo}; in our analysis the limits from Rubidium and Ytterbium clocks (lines 6 and 7) have been updated to those in lines 8 and 9.}
\end{table}

\begin{figure}i
\includegraphics[width=8cm]{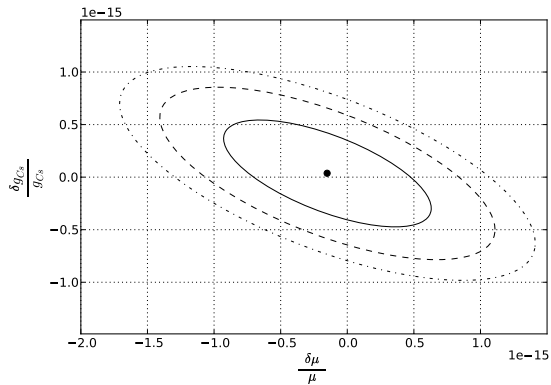}
\includegraphics[width=8cm]{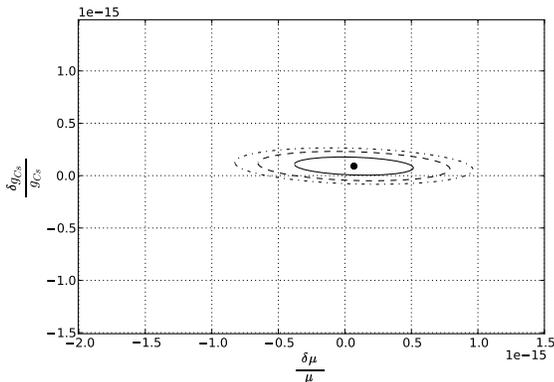}
\vskip-1cm
\caption{\label{fig1} Atomic clock constraints on the $\mu$-$g_{Cs}$ parameter space. The top panel shows the constraints obtained with the data discussed in \protect\cite{Luo}, while the bottom panel shows the constraints derived from the most recent data, ie using \protect\cite{Peik2,Bize2} instead of \protect\cite{Peik1,Bize1}. In both cases the one-, two- and three-sigma likelihood contours are plotted. Notice the change in the degeneracy direction.}
\end{figure}

\begin{figure}
\includegraphics[width=8cm]{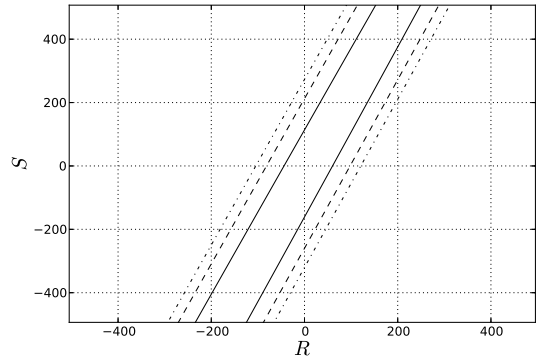}
\includegraphics[width=8cm]{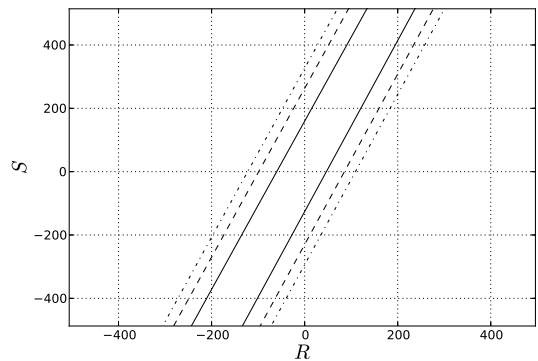}
\vskip-1cm
\caption{\label{fig2} Atomic clock constraints on the $R-S$ parameter space, using the same data as for the bottom panel of Fig. \protect\ref{fig1}. The one-, two- and three-sigma likelihood contours are plotted. In the top panel the relations between the variations of the gyromagentic rations and $\alpha$ are assumed to be those of Eqs. (\protect\ref{gyro1a}-\ref{gyro1b}), while the bottom panel assumes Eqs. (\protect\ref{gyro2a}-\ref{gyro2b}).}
\end{figure}

\section{\label{data}Constraints from atomic clocks}

By measuring the rate of two different atomic clocks one obtains a constraint on the relative shift of the corresponding characteristic frequencies. These are proportional to certain products of fundamental couplings, and thus the measurement can be translated into a constraint of the drift of that combination. Different clock comparisons are sensitive to different products of these couplings, and therefore a combined analysis can in principle lead to constraints on each of them.

Table \ref{table1} shows the existing constraints for several pairs of clocks. Since the Hg-Al comparison yields a direct constraint on $\alpha$ \cite{Rosenband}, one can use the combined dataset to obtain constraints in the $\mu$-$g_{Cs}$ plane. In \cite{Luo} this was done for the first 7 entries on the table, and we reproduce this (for comparison purposes) in the top panel of Fig. \ref{fig1}. The bounds coming from Rubidium and Ytterbium clocks (lines 6 and 7) have since improved to those in lines 8 and 9, and a reanalysis leads to the improved constraints in the bottom panel of Fig. \ref{fig1}.

Notice that the new measurements must replace the old ones in the analysis, as they are not independent---in both cases, the improved results are primarily due to a longer comparison time for the same set of clocks. With the more recent measurements the degeneracy direction is significantly changed. This is due to the fact that the Rubidium measurement (which is now the most sensitive one, apart from the $\alpha$-only one) is not sensitive to $\mu$.

From this combined analysis we can also calculate the 95\% confidence intervals for both parameters, finding
\begin{equation}
\frac{\dot\mu}{\mu}=(6.8\pm57.6)\times10^{-17}\,{\rm yr}{}^{-1}\,,
\end{equation}
\begin{equation}
\frac{\dot g_{Cs}}{g_{Cs}}=(9.1\pm11.3)\times10^{-17}\,{\rm yr}{}^{-1}\,;
\end{equation}
the latter can be equivalently expressed in terms of $g_p$
\begin{equation}
\frac{\dot g_p}{g_p}=(-7.2\pm8.9)\times10^{-17}\,{\rm yr}{}^{-1}\,.
\end{equation}
These should be compared to the result of \cite{Rosenband} for the fine-structure constant (also at the 95\% confidence level)
\begin{equation}
\frac{\dot \alpha}{\alpha}=(-1.7\pm4.9)\times10^{-17}\,{\rm yr}{}^{-1}\,.
\end{equation}
There is no evidence of variations. The bound for $g_p$ is almost as strong the one for $\alpha$, whereas the one for $\mu$ is significantly weaker. This highlights the importance of improved experimental bounds using pairs of clocks with different sensitivities to $\alpha$, $\mu$ and $g_p$.

The formalism in the previous section can be used to obtain constraints on the R-S parameter space, shown in Fig. \ref{fig2}. As we pointed out, the relations between the gyromagnetic rations and $\alpha$ are given by Eqs. (\protect\ref{gyro1a}-\ref{gyro1b}) for a simple shell model, while a better phenomenological description yields Eqs. (\protect\ref{gyro2a}-\ref{gyro2b}). Fig. \ref{fig2} presents the results for both assumptions, quantifying the importance of this theoretical uncertainty: with current experimental data this is not critical, but as measurements improve better theoretical calculations will become necessary.

Here there is a degeneracy between the two parameters, so that only a combination of them can be reasonably well constrained. The degeneracy direction can be characterized by $(S+1)-2.7R=-5\pm15$, and the allowed region has a relatively large uncertainty due to the fact that $g_p$ is less sensitive than $\mu$ to the values of $R$ and $S$. The naively expected values (for typical GUT models) of $R\sim30$, $S\sim160$ \cite{Coc,Luo} are allowed by the experimental results. By separately fixing each of them we find the following conservative bounds for the other
\begin{equation}
R=61\pm71\,,\quad{\rm assuming}\, S=160
\end{equation}
\begin{equation}
S=76\pm197\,,\quad{\rm assuming}\, R=30\,.
\end{equation}
These values are in agreement, at the 95\% confidence, with both methods of calculation depicted in Fig. \ref{fig2}.

\section{\label{concl}Conclusions}

We have considered the latest tests of the stability of nature's fundamental constants using atomic clocks and discussed their usage as a tool to constrain unification scenarios. A global analysis of existing measurements, assuming the tight bound of Rosenband \textit{et al.} \cite{Rosenband}, allows us to obtain separate updated constraints on $\mu$ and $g_p$.

It's worth comparing our results with the ones recently found by \cite{Bize2}. Although they use a different parametrization that does not explicitly include $g_p$ (and also a slightly different dataset), the results of both works agree in the case of $\mu$: at two sigma \cite{Bize2} find ${\dot\mu}/{\mu}=(15\pm60)\times10^{-17}\,{\rm yr}{}^{-1}$, while we find the marginally tighter ${\dot\mu}/{\mu}=(6.8\pm57.6)\times10^{-17}\,{\rm yr}{}^{-1}$. On the other hand they find a relatively weak bound for the ratio of the quark mass to the QCD mass scale, while we find a comparatively stronger bound for $g_p$; the difference is due to the fact that atomic clock experiments are more sensitive to $g_p$ than to the ratio $m_q/\Lambda_{QCD}$.

We carried out a first analysis of the impact of atomic clock measurements on the phenomenological parameters describing the class of varying fundamental coupling scenarios under consideration: $R$, related to QCD physics, and $S$, related to electroweak/Higgs physics. These measurements are only sensitive to a particular combination of these parameters. The experimental results are in agreement with theoretical expectations on unification scenarios.

This R-S degeneracy can be broken by measurements in astrophysical systems that have different sensitivities to these parameters. Two examples of such systems are main sequence stars and neutron stars, for which parts of the R-S parameter space have been previously explored in \cite{Vieira,Angeles}. We will discuss these issues in a future publication.

\begin{acknowledgments} 
This work was done in the context of the project PTDC/FIS/111725/2009 from FCT (Portugal), with additional support from grant PP-IJUP2011-212 (funded by U. Porto and Santander-Totta). The work of CJM is supported by a Ci\^encia2007 Research Contract, funded by FCT/MCTES (Portugal) and POPH/FSE (EC).

We are grateful to Nelson Nunes and Jean-Philippe Uzan for their comments and suggestions.
\end{acknowledgments}

\bibliography{joint}

\end{document}